# Heterogeneous nucleation of catalyst-free InAs nanowires on silicon

U P Gomes[1], D Ercolani[1], V Zannier[1], S Battiato[1], E Ubyivovk[2,3], V Mikhailovskii[2], Y Murata[1], S Heun[1], F Beltram[1] and L Sorba[1]

[1] NEST, Scuola Normale Superiore and Istituto Nanoscienze-CNR, I-56127 Pisa, Italy
[2] St. Petersburg State University, St. Petersburg, 198504 Russia
[3] ITMO University, 49 Kronverksky Pr., St. Petersburg 197101, Russia

email: daniele.ercolani@sns.it

**Abstract.** We report on the heterogeneous nucleation of catalyst-free InAs nanowires on Si (111) substrates by chemical beam epitaxy. We show that nanowire nucleation is enhanced by sputtering the silicon substrate with energetic particles. We argue that particle bombardment introduces lattice defects on the silicon surface that serve as preferential nucleation sites. The formation of these nucleation sites can be controlled by the sputtering parameters, allowing the control of nanowire density in a wide range. Nanowire nucleation is accompanied by unwanted parasitic islands, but by careful choice of annealing and growth temperature allows to strongly reduce the relative density of these islands and to realize samples with high nanowire yield.

## 1. Introduction

Monolithic integration of III-V nanowires on silicon substrates has gained considerable research interest because of its significant potential for future electronic and opto-electronic devices. Indeed, a broad range of devices like light emitting diodes (LEDs) [1], lasers [2], field effect transistors (FET) [3], and solar cells [4] were already demonstrated based on III-V NWs grown on silicon substrates. Interesting device applications like tunnel diodes and tunnel FETs [5] were also reported by utilizing III-V/silicon heterojunctions.

Several silicon substrate preparation approaches were employed for reproducible and controllable NW growth. Self-assembled III-V NWs were grown on oxide-masked silicon with nanometer-sized pin-holes [9], while site-controlled growth was demonstrated using selective area epitaxy (SAE) [10]. NW density and morphology were shown to be controlled by the oxide mask itself, but this in turn requires precise control over the chemical and physical properties of the oxide [11] thus making this approach rather time consuming and challenging contrary to simple case when NWs are directly grown on silicon substrates[12]. Actually, direct growth of III-V NWs, in particular InAs on silicon, does have critical issues. InAs NW growth is usually accompanied by unwanted parasitic island formation and both continue to nucleate even after long growth

The common way of to grow NWs is the VLS mechanism in which foreign metal nanoparticles deposited on the silicon substrate catalyse the growth [6]. Catalyst-free growth techniques are however intensely explored driven by the need to avoid metal contamination and favour its full compatibility with the present silicon technology[7,8].

times [13,14]. These islands cover a substantial fraction of the silicon surface, which is further decreased by additional nucleation leading to coalescence of NWs and islands. Not only does coalescence reduce the density of freestanding NWs, it also results in the deterioration of the NW morphological, structural, and optical properties [15]. For most practical purposes, it is desirable to suppress the coalescence process by reducing the initial nucleation density as well as inhibiting new nucleation. However, there are very few reports on density-control methodologies for InAs NWs on silicon [16,17]. Therefore, there is an urgent need for understanding the nucleation mechanisms of NWs and islands deposited on silicon to achieve control on their density.

In this work, we present a new approach to prepare Si (111) surfaces for the controlled growth of InAs NWs by chemical beam epitaxy (CBE). We show that controlled sputtering of the surface of Si (111) enhances the nucleation of InAs crystals (NWs and islands) while no nucleation occurs on non-sputtered Si (111) surfaces under identical growth conditions. We argue that this stems from the formation of surface defects associated to the sputtering process that serve as preferential physical nucleation sites. Furthermore, we show that the silicon surface can be modified by *in situ* growth and *ex situ* sputtering parameters allowing to control the density of InAs NWs over a wide range. We show that the yield of NWs with respect to islands can be maximized by choosing an optimum growth window obtaining InAs NW densities in the range of ~1-30 NWs/µm$^2$ with a yield of ~50%.

## 2. Experimental details

The substrate preparation protocol is schematized in figure 1. Commercially available Si (111) substrates covered with a native or 20 nm thick thermal-oxide layer were used (figure 1(a)). The oxide was completely removed by a 2 minute buffered oxide etch (BOE) having an etch rate of ~90 nm/min. The substrates were then rinsed in deionized water and blown dry with nitrogen to obtain hydrogen terminated Si (111) surfaces. These substrates will be termed "non-sputtered" silicon substrates in the following (figure 1(b)). No NW growth occurred on non-sputtered silicon substrates (figure 1(c)), and hence were used for further surface treatments.

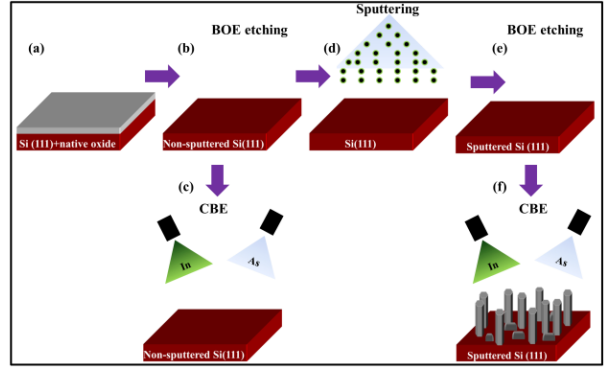

**Figure 1** Schematic of the processes used for substrate preparation of non-sputtered silicon (a-c) and of sputtered silicon (a-f).

Subsequent surface treatment involved sputtering the silicon surface by controlled Argon or $SiO_2$ energetic particles (figure 1(d)). Energetic Argon particles were obtained by an inductively coupled plasma (ICP) sputtering system (chamber pressure= 1.5 ×10$^{-4}$ Torr, Argon flow=15 sccm), while $SiO_2$ particles are obtained by a radio-frequency (RF) magnetron sputtering system (chamber pressure= 1×10$^{-4}$ Torr). After sputtering and prior to NW growth, the sputtered substrates were again etched for 2 minutes in BOE to remove residuals and deposited oxide. The substrates were rinsed in deionized water and blown dry with nitrogen. These substrates will be termed "sputtered" silicon substrates (figure 1(e)) and used for NW growth (figure 1(f)).

Immediately after etching, all substrates were indium-bonded on a molybdenum platen in air by keeping them on a hot plate at 250°C for less than a minute, transferred to the load lock of the CBE chamber and pumped to 10$^{-7}$ Torr in few minutes. Auger electron spectroscopy of "as mounted" substrates reveals that some silicon oxides are present on the surface (see supplementary information S1, figure S1), as a consequence of the indium bonding procedure. Although the amount of oxide is very small, it indicates that substrate handling before mounting requires particular care.

The process employed for growing InAs NWs involved a high temperature annealing at a temperature $T_{ann}$ under As flux corresponding to a TBAs line pressure of 1.0 Torr for 15 minutes. After this step, the temperature was ramped down to the growth temperature $T_{growth}$, and InAs growth was initiated by introducing As and In fluxes at metal-organic line pressures $P_{TBAs}$=3.0 Torr and $P_{TMIn}$=0.3 Torr, respectively for a growth time $t_{growth}$. The NW growth was terminated by cooling down the samples under As flux. All temperatures were measured using a pyrometer with an accuracy of ±10°C.

Field-emission scanning electron microscopy (SEM) images of the NWs were acquired with a Zeiss Merlin SEM operating typically at 5 keV. For each sample, the density was measured from plan-view SEM micrographs taken from at least about 10 random areas of the sample. The minimum measurable InAs crystal size is ~ 3 nm (SEM image resolution). The statistical data on the density were obtained using open-source Image J software by automatic particle counting after thresholding the SEM images. High-resolution transmission electron micrographs (HRTEM) were acquired using a Libra 200 TEM operating at an accelerating voltage of 200 kV. The incident electron beam was in the vicinity of <110> silicon zone axis. The samples were sliced into thin sections by focused-ion-beam (FIB).

### 3. Nucleation and growth mechanism

Figure 2 shows plan-view SEM micrographs of InAs NWs grown on non-sputtered Si (111) (figure 2(a)) and on sputtered Si (111) substrates employing two different sputtering procedures: $SiO_2$ sputtering using RF magnetron sputtering system (figure 2(b)) and Argon sputtering using an ICP system (figure 2(c)). The sputtered surface using $SiO_2$ sputtering was obtained using a sputter bias ($V_{sputter}$) of 520 V and a sputter time ($t_{sputter}$) of 60 s. $V_{sputter}$ and $t_{sputter}$ were 35 V and 600 s, respectively, for ICP Argon sputtering. The NW growth process involved annealing at $T_{ann}$=790 °C and the growth at $T_{growth}$=400 °C for $t_{growth}$ =15 min.

SEM micrographs in figures 2(a-c) show that InAs crystals nucleate only on the sputtered Si (111) surfaces. Extensive investigation of non-sputtered samples grown with a wide variety of growth parameters never revealed NWs or islands except near borders or recognizable defected areas of the substrate. The enhancement of nucleation on sputtered Si (111) surfaces in comparison to that of non-sputtered Si (111) surfaces is therefore likely due to the formation of surface defects by sputtering [18]. Although a large number of processes occur during sputtering, it is well established that bomb-arding a surface with energetic particles leads to the formation of surface defects, which here serve as preferential nucleation sites [18,19]. In fact, the role of surface defects as nucleation sites in catalyst-free NW growth is already established: it was found that defects ease the formation of critical nuclei and enhance the nucleation rate [20,21].

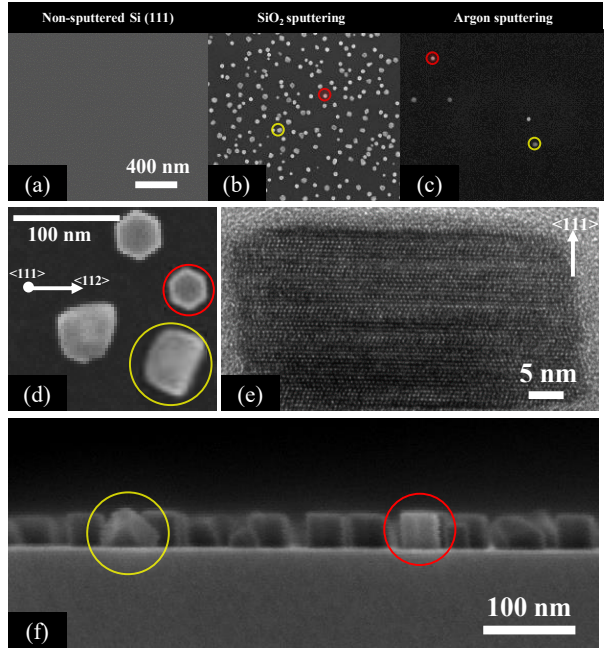

**Figure 2** Plan-view SEM micrographs of InAs crystals grown under identical conditions on (a) non-sputtered, (b) $SiO_2$ sputtered, and (c) ICP Argon sputtered silicon substrates. (d) Plan-view SEM micrograph of InAs NWs and islands from the sample shown in (b). The coloured markers are used to identify islands (yellow) and NWs (red). (e) High-resolution TEM image of an individual InAs NW from the sample shown in (b) taken in the Si <110> zone axis. (f) 90° tilted SEM micrograph of the same sample as in (b).

The different InAs crystal density on the two types of sputtered silicon surfaces shown in figures 2(b) and (c) is likely due to the different sputtering parameters ($V_{sputter}$ and $t_{sputter}$) and techniques employed. The details of the influence of sputtering parameters on the InAs crystal density ($N$), which is the sum of InAs NW and island density, for the two sputtering techniques are discussed in the supplementary information (see figures S2 and S3). It is shown that $N$ increases with $V_{sputter}$ and $t_{sputter}$ as long as surface defects favourable for nucleation are created by the sputtering particles. Although $N$ depends strongly on $V_{sputter}$ and $t_{sputter}$ (see supplementary information S2 and S3), a direct comparison of the two techniques is not straightforward. This is because the impact of the particles impinging on the surface depends not only on $V_{sputter}$ but also on the mass of the sputtering species, chamber pressure and geometry, which are different in ICP Argon and RF magnetron $SiO_2$ sputtering systems. The chamber pressure and geometry can influence the mean free path of the sputtering particles and thus their energies. In addition, an oxide layer is deposited during $SiO_2$ sputtering that can inhibit the creation of new defects when the thickness of the deposited layer is greater than the penetration depth of the sputtering particles.

The InAs crystals shown in figures 2(b) and (c) consist of NWs (red circles) and islands (yellow circles). Figure 2(d) shows a magnified view of InAs NWs and islands and highlights their morphological differences. NWs, (red circles) have a hexagonal cross-section delimited by six {112} facets normal to the Si(111) surface. On the other hand, islands (yellow circles), display irregular cross-sections. For a given sample, the yield ($Y$) of NWs is determined as the ratio of NW density to the total crystal (NWs and islands) density. It is ~0.5 for the samples shown in figures 2(b) and (c). As will be shown later, $Y$ can be controlled by the annealing and growth conditions. Figure 2(e) shows a cross-sectional HRTEM image of a NW taken from the sample shown in figure 2(b). The micrograph is taken in the Si <110> zone axis with the NW growing along the vertical [111] axis. At this early stage, for the chosen growth parameters, the NWs have aspect ratio less than one and display vertical {112} sidewalls. NWs are growing epitaxially on the substrate, with horizontal InAs (111) planes and a flat top facet. Already at this early stage of growth, the crystal structure has a high stacking fault density showing intermixing of wurtzite and zincblende layers. Overall, the morphology is equivalent to that of longer InAs NWs previously reported [7]. Figure 2(f) is a 90° tilted SEM micrograph of the sample of figure 2(b) with the electron beam impinging normally on a Si {110} cleavage plane, showing again the <111> oriented growth of the NWs. As in the other panels, the red and yellow circles highlight one NW and one island, respectively.

In order to identify defect formation on the sputtered silicon surface, we investigated the characteristic features after sputtering. It is well known that sputtering can transform a crystalline material into an amorphous state when the density of defects becomes very high [22]. However, the NW growth with the <111> axis normal to the Si surface (see figure 2(f)) suggests that the crystalline order of the silicon substrate surface is not destroyed by sputtering and that the surface is not amorphized. This is confirmed by HRTEM analysis, which shows defect-free crystalline order of the sputtered Si substrates after growth.  An example of micrograph acquired in the Si <110> zone axis is displayed in figure 3(a).  The darker region in the bottom of the image is the Si substrate, where the crystal lattice is evident. The darker areas near the NW base are due to strain contrast derived from the large lattice mismatch between Si and InAs. Higher energy sputtering ($V_{sputter}$ > 40 V in ICP Argon sputtering and $V_{sputter}$ > 520 V in RF magnetron $SiO_2$ sputtering) results in a decrease in the InAs crystal density (see figures S2 and S3 in supplementary information). Hence, we argue that the nucleation sites that mediate epitaxial growth may be surface defects formed at low values of $V_{sputter}$ while higher energy particles, with greater penetration depth, etch and/or amorphize the silicon substrate. Therefore proper care must be taken to control the energy of the sputtered particles so as not to etch or amorphize the silicon surface.

As for sputtering-induced surface roughness, it is unlikely to be the source of nucleation sites [23,24]. We have analysed it from AFM surface topographies, without observing substantial differences between the non-sputtered and sputtered surfaces. The roughness has been determined by AFM scans on freshly BOE-etched substrates, using various resolutions and scan sizes. After a surface flattening to remove acquisition artefacts, the roughness has been calculated as the average roughness (Ra) using WsXM software [25]. Depending on the individual scan, Ra values show some variability.  The mean Ra from all the scans (±standard deviation) is 0.38±0.19 nm and 0.59±0.29 nm for the non-sputtered and sputtered ($SiO_2$ sputtering, $V_{sputter}$ =520 V, $t_{sputter}$=60 s) surfaces, respectively.  Though a slight difference is observable, it is well within the uncertainties of the measured roughness. Representative AFM topographs are reported in figures 3(b) and (c) for a non-sputtered and a sputtered sample, respectively.

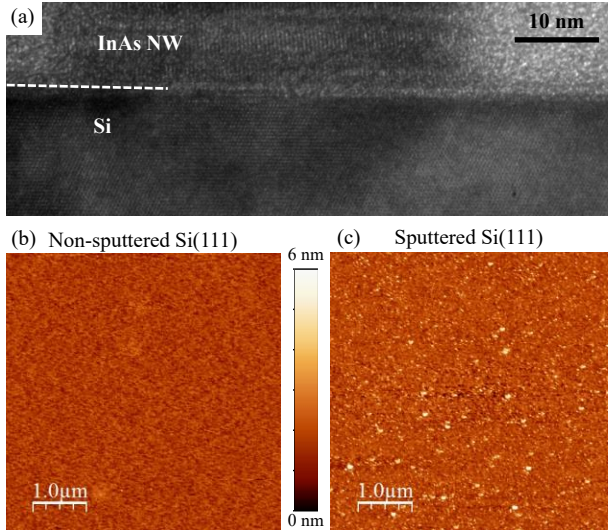

**Figure 3** (a) TEM micrograph of the Si surface acquired in the Si <110> zone axis of a sputtered sample after NW growth. The dashed line marks the Si {111} surface. A defect-free Si lattice is clearly resolved. AFM topographies of (b) non-sputtered and (c) sputtered Si surface, displaying similar roughness.

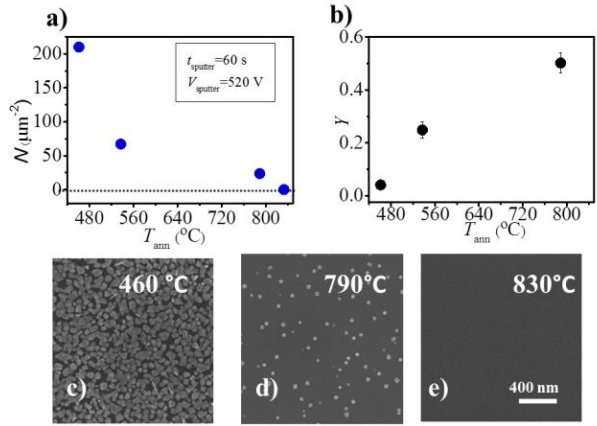

**Figure 4** Correlation between (a) InAs crystal density ($N$) and (b) NW yield ($Y$) with annealing temperature. (c)-(e) selected plan-view SEM micrographs of samples annealed at 460, 790 and 830 °C, respectively.

The final confirmation of the role of surface defects as InAs nucleation centers is shown in figure 4. Figures 4(a) and (b) show the dependence of $N$ and $Y$, respectively, as a function of $T_{ann}$ for InAs crystals grown on SiO$_2$ sputtered silicon substrates ($V_{sputter}$ =520 V, $t_{sputter}$=60 s). Figures 4(c)-(e) are selected SEM micrographs of samples grown after annealing at different $T_{ann}$. Annealing time was kept fixed at 15 minutes and the process was performed in an As flux corresponding to $P_{TBAs}$ of 1.0 Torr. Growth was carried out at $T_{growth}$=400 °C for 15 minutes.

Figure 4(a) demonstrates an appreciable reduction of InAs crystal density ($N$) with an increase in the annealing temperature. At the lowest $T_{ann}$ of 450 °C, $N$ is highest (~210 crystals/µm$^2$) while at the highest $T_{ann}$ of 830 °C, InAs crystals are completely absent. This reduction in $N$ observed due to annealing suggests that more surface defects annihilate as $T_{ann}$ is increased and at the highest $T_{ann}$ of 830 °C the sputtered silicon surface regenerates into an ordered silicon surface, comparable to a non-sputtered silicon surface (see figure 2(a)).

Figure 4(b) shows that the yield of NWs ($Y$) increases with $T_{ann}$, and at $T_{ann}$ =790 °C, $Y$ is maximum (~ 0.5). At $T_{ann}$ > 790 °C there is no InAs crystal nucleation (see figure 4(a)) and hence $Y$ is undefined.

The nucleation mechanism of islands and NWs depends on the nature of the Si (111) surface. Annealing and cooling cycles under UHV conditions not only determine the defect density but also the surface reconstruction [26,27]. It is known that different surface reconstructions have different surface energies that determine the polarity and morphology of InAs crystals [28–30]. It is likely that the surface reconstructions formed below $T_{ann}$ =790 °C do not favour NW formation, and therefore the yield is low. On the other hand, we expect that the annealing process at $T_{ann}$ =790 °C for 15 minutes and the subsequent cool down to the growth temperature of 400 °C in As background transforms a large portion of the surface from Si (1×1)-H to Si (1×1)-As [27,31]. At annealing temperature up to approximately 500 °C hydrogen atoms desorb from the hydrogen passivated silicon surface and becomes As-passivated in As ambient [29]. An As-passivated silicon surface was found to have a very low surface energy due to the existence of As lone-pair states inducing Volmer-Weber (VW) growth of InAs islands [32]. Moreover, the As-passivated silicon surface was shown to resemble an InAs (111)B oriented surface and nucleation of VW islands on such a surface resulted in a high yield of vertically oriented catalyst-free NWs [27,29,33]. Although the As-passivated silicon surface should be very stable [32], the passivation of the surface may still be incomplete due to the temperature dependence of adsorption and desorption of As atoms on silicon [31]. The silicon surface may therefore contain at the same time some portions of unpassivated surface [34] and surfaces that resemble the InAs (111)A surface [27,35]. These surfaces have different surface energies in comparison to the As-passivated silicon surface. The VW islands that nucleate on unpassivated surfaces and surfaces that resemble the InAs (111)A surface may not develop the top (111) facets and therefore continue to grow isotopically as islands as reported in [29,35,36]. This simple argument applies also to our case where 100% yield of NWs was not achieved. Overall, these results indicate that a spectrum of defect sites exists on the sputtered substrate enabling InAs crystal nucleation whose density and morphology depends on the annealing conditions.

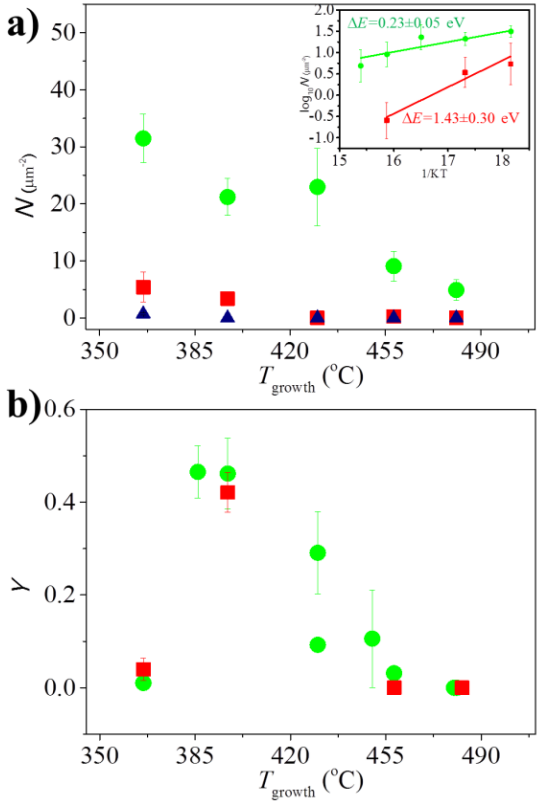

**Figure 5** Growth temperature dependence of InAs crystal density ($N$) for three different substrate preparation protocols using $SiO_2$ sputtering with sputtering parameters $t_{sputter}$=60 s, $V_{sputter}$=520 V (green circles) and $t_{sputter}$=3 s, $V_{sputter}$=345 V (red squares) and ICP Argon sputtering with sputtering parameters $t_{sputter}$=600 s, $V_{sputter}$=35 V (blue triangles). The inset shows an Arrhenius plot of the same data for the $SiO_2$ sputtered samples from which the activation energy is determined. (b) Yield of InAs NWs as a function of $T_{growth}$ for the $SiO_2$ sputtered samples shown in figure 5(a).

In order to investigate the nucleation process further, we analyse the InAs crystal density ($N$) as a function of growth temperature. Figure 5(a) shows the variation of $N$ as a function of $T_{growth}$ for SiO2 sputtered silicon substrates employing two different sputtering parameters, and for an ICP Argon sputtered silicon surface. NW growth time was set at 15 minutes and was preceded by annealing at $T_{ann}$=790 °C for 15 minutes. The inset shows an Arrhenius plot of the same set of data from $SiO_2$ sputtered silicon substrates from which the activation energy is derived. Figure 5(b) shows the variation of $Y$ as a function of $T_{growth}$ on the $SiO_2$ sputtered substrates.

Figure 5(a) reveals that $N$ strongly decreases with increasing growth temperature for all types of sputtered substrates. The crystal density $N$ for the substrate sputtered with $SiO_2$ with parameters $t_{sputter}$=3 s and $V_{sputter}$=345 V (red squares) is reduced to zero at 420 °C while at the same temperature $N$=25/μm$^2$ for the substrate with $SiO_2$ sputtering parameters $t_{sputter}$=60 s and $V_{sputter}$=520 V (green circles). On the other hand, the ICP Argon sputtered substrates have a very narrow temperature window with $N$=0.7/μm$^2$ at 365 °C (blue triangles), while no nucleation takes place above 400 °C for a fixed growth time of 15 minutes.

This simple observation can be rationalized by the island nucleation theory [19,37,38]. According to the theory, the crystal density, $N$, is given as [38]:

$$N \propto N_D \exp\left(-\Delta E_b / KT_{growth}\right) \exp\left(\Delta E_T / KT_{growth}\right) \quad (1)$$

here $N_D$ is the density of surface defects and $\Delta E_b$ is the nucleation energy barrier associated with defect-mediated heterogeneous nucleation. $\Delta E_T$ is the sum of a number of energy terms for thermally activated processes accounting for surface diffusion, desorption, dissociation of InAs critical nuclei, etc. [37,39]. Generally, a high density of defects effectively reduces $\Delta E_b$ and enhances the nucleation rate.

According to Eq. 1, the first exponential will lead to an increase and the second in a decrease of the nucleation density with $T_{growth}$. Therefore, the decrease in $N$ with increasing $T_{growth}$ as shown in figure 5(a) can be linked to the increase in thermally activated processes like surface diffusion and desorption. The total activation energy, $\Delta E$ is the sum of $\Delta E_b$ and $\Delta E_T$. $\Delta E$ is derived from the Arrhenius plot (see inset of figure 5(a)) and is 0.23±0.05 eV for SiO$_2$ sputtered silicon with parameters $t_{sputter}$=60 s at 520 V and 1.43±0.30 eV for sputtered silicon with parameters $t_{sputter}$=3 s at 345 V. These results confirm that increased sputtering time and voltage increase the surface defect density and lowers the total nucleation energy barrier when grown under identical conditions. As $N$ is 0 for ICP Argon sputtered silicon at temperature greater than 400 °C, no Arrhenius plot could be obtained.

It can be seen in figure 5(b) that there is an optimum temperature window in which the yield of NWs is maximum. It is seen that within the temperature window of 385-400 °C, $Y$~0.5, i.e. 50% of the InAs crystals nucleate as NWs. In order to understand this, the island nucleation theory must be coupled with the self-induced nucleation mechanism. According to the self-induced nucleation mechanism, NWs are formed from three dimensional VW islands after undergoing a series of shape transformations [8,40]. Such a process is thermally activated and an optimum temperature is required to overcome the energy barrier for shape transformations [41–43]. At this optimum temperature, a large fraction of three-dimensional VW islands can transform into NWs thereby increasing the yield as shown in figure 5(b). Furthermore, the yield of NWs is independent of the sputtering parameters ($t_{sputter}$ and $V_{sputter}$) and is ~0.5 for all samples annealed at $T_{ann}$=790 °C for 15 minutes and grown for 15 minutes at 400 °C (see figure S2(c) in supplementary information).

The nucleation mechanism was also investigated as a function of growth time, $t_{growth}$. Figures 6(a) and (b) report $N$ as a function of $t_{growth}$ for samples grown on ICP Argon sputtered substrates with sputtering parameters $t_{sputter}$=600 s, $V_{sputter}$=35 V and on SiO$_2$ sputtered substrates with sputtering parameters $t_{sputter}$=60 s and $V_{sputter}$=520 V, respectively . Figure 6(c) shows a SEM micrograph of a sample grown for 135 minutes on an ICP Argon sputtered substrate. Figures 6(d) and (e) show SEM micrographs of samples grown for 15 and 60 minutes, respectively, on SiO$_2$ sputtered substrates. Growth is conducted at $T_{growth}$=400 °C.

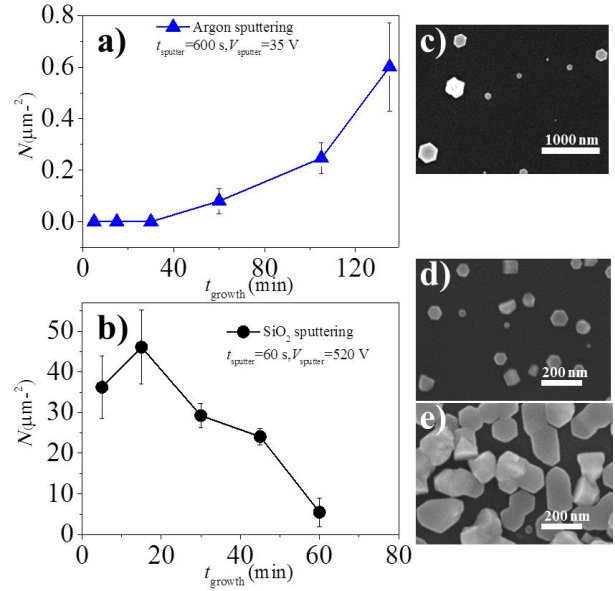

**Figure 6** InAs crystal density ($N$) as a function of growth time on (a) ICP Argon sputtered substrates and (b) on SiO$_2$ sputtered substrates. Plan-view SEM micrograph of InAs crystals (c) after 135 minutes of growth on ICP Argon sputtered substrate, (d) after 15 minutes and (e) after 60 minutes of growth on SiO$_2$ sputtered substrate.

For the ICP Argon sputtered substrates, as shown in figure 6(a), $N$ increases with increasing $t_{growth}$ due to continuous new nucleation. Such a process causes a large NW diameter distribution as shown for a sample grown for 135 minutes in figure 6(c).

A large distribution of NW diameters is also evident in SiO$_2$ sputtered substrates as seen in figure 6(d) owing to new nucleation of InAs crystals. However, no significant increase in $N$ with $t_{growth}$ is seen in SiO$_2$ sputtered substrates (figure 6(b)). This is because the variation of $N$ with $t_{growth}$ is strongly affected by (i) surface coverage and (ii) coalescence effects. The high density of InAs crystals nucleated at the early stages of growth covers a large fraction of the silicon surface. As the growth time is increased, the size of the crystals increases which further increase the surface coverage. High surface coverage reduces new nucleation events so that $N$ does not change appreciably [19,44]. Furthermore, the coalescence of InAs crystals reduces the overall crystal density, and hence the occurrence of new nucleation cannot be reliably inferred from figure 6(b). The coalescence can be seen in plan-view SEM micrograph in figure 6(e) for a sample grown for 60 minutes.

## 4. Conclusions

In conclusion, we have shown that InAs crystals nucleate on sputtered Si (111) surfaces while no nucleation occurs on non-sputtered Si (111) surfaces. We discuss the role of surface defects as preferential nucleation sites formed by sputtering under controlled parameters using different sputtering techniques. The InAs crystals nucleating on the sputtered silicon surfaces consist of InAs NWs and parasitic islands. Although the nucleation of parasitic islands could not be completely inhibited, the yield of NWs could be increased up to ~0.5 by proper choice of growth and annealing temperatures. We show that a range of InAs NW densities can be obtained by modifying *in situ* growth and *ex situ* sputtering parameters.

## Acknowledgments

L.S, S.B, V.Z and U.G gratefully acknowledge the project CNR-RFBR. The TEM lamellae preparation and TEM data was obtained using the equipment of the Interdisciplinary Resource Centre for Nanotechnology of St. Petersburg State University, Russia. E.U and V.M acknowledges RFBR (grant No.15-02-06525) and St. Petersburg State University (Project No. 11.37.210.2016) for the financial support.

# Supplementary Information for

# Heterogeneous nucleation of catalyst-free InAs nanowires on silicon

U. P. Gomes[1], D. Ercolani[1], V. Zannier[1], S. Battiato[1], E. Ubyivovk[2,3], V. Mikhailovskii[2], Y. Murata[1], S. Heun[1], F. Beltram[1] and L. Sorba[1]

1. NEST, Scuola Normale Superiore and Istituto di Nanoscienze-CNR, I-56127 Pisa, Italy
2. St. Petersburg State University, St. Petersburg, 198504 Russia
3. ITMO University, 49 Kronverksky Pr., St. Petersburg 197101, Russia.

**S1-Auger Analysis**

Figure S1 shows three Auger spectra obtained from a Si (111) surface covered with a thin layer of native oxide (black curve), a Si (111) substrate obtained after etching the native oxide by BOE (green curve) and a sputtered Si (111) substrate (red curve). The sputtered Si (111) substrate is obtained by RF magnetron $SiO_2$ sputtering ($V_{sputter}$ =520 V, $t_{sputter}$=60 s) followed by etching for 2 minutes in BOE and then placing it on a hotplate at 250 °C for 1 minute. After BOE etching, the substrates are rinsed in deionized water and blown dry with nitrogen. Auger spectra were measured at a primary electron energy of 2 kV.

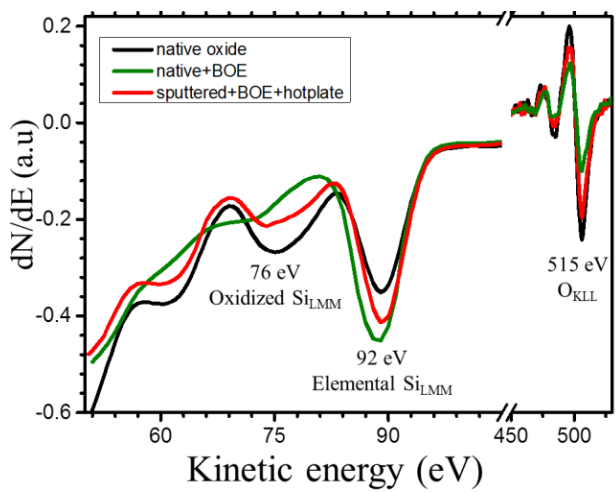

Figure S1 Auger spectra acquired from Si (111) with native oxide (black curve), after BOE etching (green curve), and from a $SiO_2$ sputtered Si (111) surface after etching and placing it on a hotplate for 1 minute (red curve).

Auger spectra of a Si (111) substrate covered with a layer of native oxide (black curve) show a peak at ~92 eV which is the elemental silicon peak. A distinct peak at ~76 eV is attributed to the chemically shifted silicon peak from $SiO_2$ [1,2]. Non-stoichiometric oxides (if present) display intermediate shifts. The structure at ~515 eV is assigned to oxygen, and stems not only from oxides but also from adsorbed water and OH groups [3].

After etching the native oxide in BOE for 2 minutes (green curve), the ~76 eV peak is almost completely suppressed confirming the complete removal of the native oxide. The oxygen peak at ~515 eV is also much smaller than that of Si (111) with native oxide. The residual oxygen detected is mainly due to residual water and OH groups [3].

When the sputtered and etched sample is placed on the hotplate for 1 minute (a step that is used to indium bond the sample on the molybdenum platen before mounting it in the CBE load-lock), a weak signal of oxidized silicon at ~72 eV appears again (red curve). The oxygen peak at ~ 515 eV is also larger than that of Si (111) substrate without native oxide but weaker than from a Si (111) substrate with a native oxide. It is likely that placing the sample on a hotplate at 250 °C in air the silicon surface re-oxidizes slightly.

## S2- RF magnetron SiO$_2$ sputtering

Figure S2(a) shows a plot of crystal density ($N$) versus $t_{sputter}$ for a fixed $V_{sputter}$=345 V. Sputtering at $V_{sputter}$=345 V corresponds to a SiO$_2$ deposition rate of 1Å/s, i.e an oxide of thickness, $t_{ox}$=20 nm is deposited for $t_{sputter}$ =200 s. Figure S2(b) shows $N$ versus $V_{sputter}$ for a fixed oxide thickness $t_{ox}$=20 nm. Figure S1(c) shows a plot of NW yield ($Y$) versus $t_{sputter}$ for a fixed $V_{sputter}$=345 V and the inset shows $Y$ versus $V_{sputter}$.

After sputtering, the substrates were etched for 2 minutes in BOE to remove the deposited oxide, rinsed in deionized water and blown dry with nitrogen. On all samples, NWs were grown at 400 °C for 15 minutes with $P_{TBAs}$=3.0 Torr and $P_{TMIn}$=0.3 Torr after annealing at 790 °C for 15 minutes in $P_{TBAs}$=1.0 Torr.

It is seen from figure S2(a) that $N$ increases with increasing $t_{sputter}$ from 3 s to 200 s. However, $N$ does not increase after 200 s of sputtering. The increase of $N$ from 3 s to 200 s is likely due to the increase of defect density with sputtering time. Instead, the saturation of $N$ after 200 s is because the deposited 20 nm oxide masks the penetration of SiO$_2$ atoms to the silicon substrate inhibiting the formation of new surface defects.

In figure S2(b) $N$ increases from 85 V to 345 V while it is reduced at the highest $V_{sputter}$=820 V. $N$ increases from 85 V to 345 V due to the increase in defect density while the reduction in $N$ at the highest bias is likely due to damage of the substrate leading to amorphization. Figure S2(c) shows that the yield of NWs is independent of both parameters, i.e $t_{sputter}$ and $V_{sputter}$ and is ~0.5.

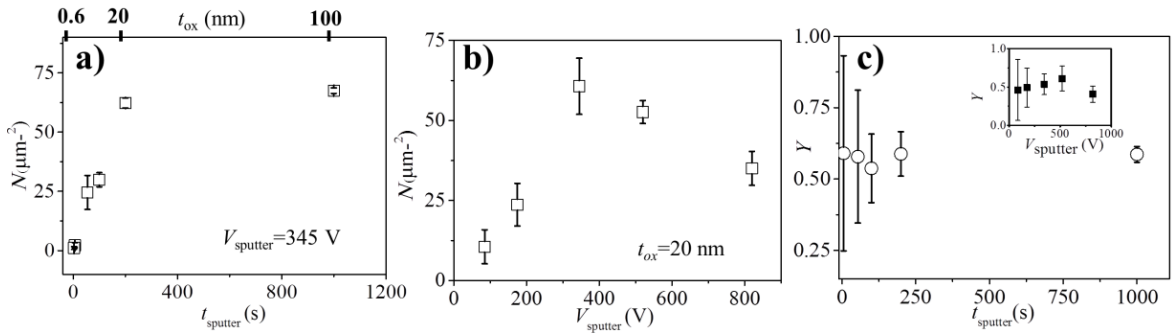

Figure S2 Variation of $N$ on SiO$_2$ sputtered silicon as a function of (a) $t_{sputter}$ and (b) $V_{sputter}$. (c) Variation of $Y$ on SiO$_2$ sputtered silicon as a function of $t_{sputter}$. The inset shows the variation of $Y$ as a function of $V_{sputter}$.

## S3-ICP Argon sputtering

Figure S3(a) shows the dependence of crystal density ($N$) on ICP Argon sputtered silicon substrate as a function of $t_{sputter}$ for a fixed $V_{sputter}$=35 V. Figure S3(b) shows a plot of $N$ vs $V_{sputter}$ for a fixed $t_{sputter}$=600 s. After sputtering, the substrates were etched for 2 minutes in BOE to remove any form of residues and oxides. The substrates were rinsed in deionized water and blown dry with nitrogen. All samples were grown at 400 °C for 15 minutes with $P_{TBAs}$=3.0 Torr and $P_{TMIn}$=0.3 Torr after annealing at 790 °C for 15 minutes in $P_{TBAs}$=1.0 Torr.

It is seen from figure S3(a) that the InAs crystal density ($N$) increases with increasing $t_{sputter}$ due to the increase of defect density with sputtering time. In figure S3(b), $N$ increases from 10 V to 40 V but reduces to zero for higher values. The reduction in $N$ at $V_{sputter}$>40 V is due to etching of the silicon surface. The yield of NWs for all the samples is in the range of 0.4-0.5.

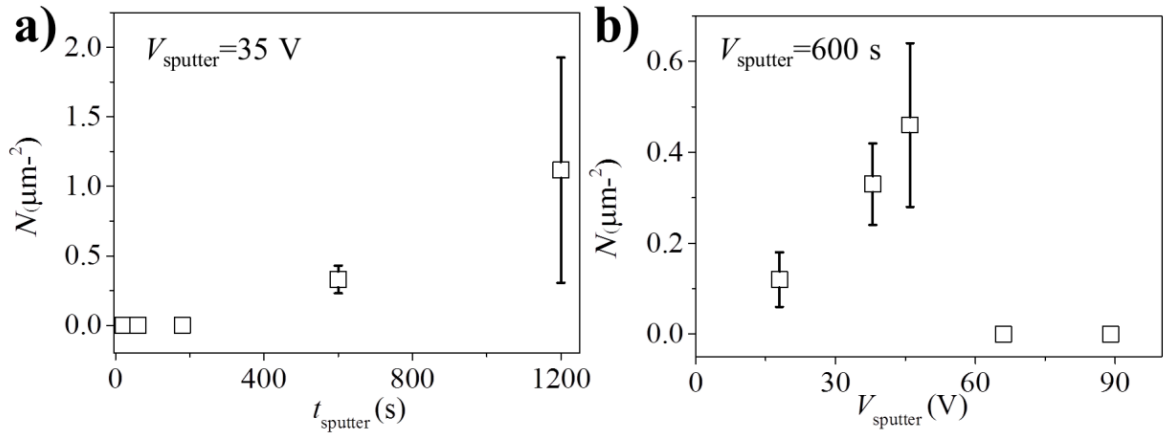

Figure S3(a) Variation of $N$ on ICP Argon sputtered silicon as a function of $t_{sputter}$ for a fixed $V_{sputter}$=35 V  (b) Variation of $N$ as a function of $V_{sputter}$ for a fixed $t_{sputter}$=600 s.